\documentclass{nature}
\usepackage[framemethod=TikZ]{mdframed}
\usepackage[T1]{fontenc}
\usepackage[latin9]{inputenc}
\usepackage{amsmath}
\usepackage{amssymb}
\usepackage{graphicx}
\usepackage{hyperref}
\usepackage[square,sort,comma,numbers]{natbib}
\usepackage{times}
\usepackage{helvet}

\makeatletter
 
 \@ifundefined{textcolor}{}
 {%
   \definecolor{BLACK}{gray}{0}
   \definecolor{WHITE}{gray}{1}
   \definecolor{RED}{rgb}{1,0,0}
   \definecolor{GREEN}{rgb}{0,1,0}
   \definecolor{BLUE}{rgb}{0,0,1}
   \definecolor{CYAN}{cmyk}{1,0,0,0}
   \definecolor{MAGENTA}{cmyk}{0,1,0,0}
   \definecolor{YELLOW}{cmyk}{0,0,1,0}
 }

\@ifundefined{textcolor}{}{%
 \definecolor{BLACK}{gray}{0}
 \definecolor{WHITE}{gray}{1}
 \definecolor{RED}{rgb}{1,0,0}
 \definecolor{GREEN}{rgb}{0,1,0}
 \definecolor{BLUE}{rgb}{0,0,1}
 \definecolor{CYAN}{cmyk}{1,0,0,0}
 \definecolor{MAGENTA}{cmyk}{0,1,0,0}
 \definecolor{YELLOW}{cmyk}{0,0,1,0}
 }

\@ifundefined{definecolor}{\@ifundefined{definecolor}
 {\@ifundefined{definecolor}
 {\@ifundefined{definecolor}
 {\@ifundefined{definecolor}
 {\@ifundefined{definecolor}
 {\usepackage{color}}{}
}{}
}{}
}{}
}{}
}{}
\makeatother

\title{Dynamics of quantum information}

\author{R.~J. Lewis-Swan$^{1,2}$, A. Safavi-Naini$^{1,2}$, A. M. Kaufman$^{1}$, and A.~M. Rey$^{1,2,3}$}

\begin{document}

\maketitle

\newcounter{theo}[section]\setcounter{theo}{0}
\renewcommand{\thetheo}{\arabic{section}.\arabic{theo}}
\newenvironment{theo}[2][]{%
\refstepcounter{theo}%
\ifstrempty{#1}%
{\mdfsetup{%
frametitle={%
\tikz[baseline=(current bounding box.east),outer sep=0pt]
\node[anchor=east,rectangle,fill=blue!20]
{\strut Box~\thetheo};}}
}%
{\mdfsetup{%
frametitle={%
\tikz[baseline=(current bounding box.east),outer sep=0pt]
\node[anchor=east,rectangle,fill=blue!20]
{\strut Box~\thetheo:~#1};}}%
}%
\mdfsetup{innertopmargin=10pt,linecolor=blue!20,%
linewidth=2pt,topline=true,%
frametitleaboveskip=\dimexpr-\ht\strutbox\relax
}
\begin{mdframed}[]\relax%
\label{#2}}{\end{mdframed}}

\begin{affiliations}
\item JILA, NIST and Department of Physics, University of Colorado, Boulder, USA
\item Center for Theory of Quantum Matter, University of Colorado, Boulder, CO 80309, USA
\item email: arey@jilau1.colorado.edu
\end{affiliations}

\begin{abstract}
The ability to harness the dynamics of quantum information and entanglement is necessary for the development of quantum technologies and the study of complex quantum systems. On the theoretical side the dynamics of quantum information is a topic that is helping us  unify and confront common problems in otherwise disparate fields in physics, such as quantum statistical mechanics and cosmology. On the experimental side the impressive developments on the manipulation of neutral atoms and trapped ions  are providing new capabilities to probe their quantum dynamics. Here, we overview and discuss progress in characterizing and understanding the dynamics of quantum entanglement and information scrambling in quantum many-body systems. The level of control attainable over both the internal and external degrees of freedom of individual particles in these systems provides great insight into the intrinsic connection between entanglement and thermodynamics, bounds on information transport and computational complexity of interacting systems. In turn this understanding should enable the realization of quantum technologies.
\end{abstract}

\section{Introduction}
One of the scientific developments in the past decade has been the emergent synergy between historically disparate fields in physics: atomic, molecular and optical (AMO) physics, condensed matter, general relativity  and high energy physics. Quantum entanglement connects these different disciplines (see Fig. 1) opening new perspectives on understanding and describing complex quantum many-body systems. A major advance is the understanding that entangled states are not only a fundamental resource for quantum information processing, but also play a crucial role in black hole thermodynamics,including  the description of a black hole's horizon \cite{Shenker2014,Hayden2007,Sekino2008,Shenker2015} and emergent space-time \cite{Maldacena2003,Ryu2006,Qi2018}. Entangled states also appear in the non-equilibrium dynamics of isolated quantum many-body systems \cite{Rigol_chaos_2016,Nandkishore_MBL_2015}. Moreover, quantum chaos is now believed to be intrinsically related to how entanglement is distributed in a system \cite{Maldacena2016,Hosur2016}; this connection has opened up a new way to define quantum chaos and tie it to complexity theory and even quantum gravity \cite{Susskind2015,Qi2018}.

Studying the dynamics of entanglement is thus important for different areas. However, it is a challenging task because entangled many-body systems are hard to create and characterize experimentally and model theoretically. Although we often understand the individual quantum building blocks well, systems involving even a few tens of quantum particles interacting with each other exhibit complex and new behaviors which are typically inaccessible to classical computer simulations. But thanks to experimental advances in controlling and manipulating atomic systems such as quantum gas microscopes, optical tweezers and arrays of trapped ions (See Fig. 1), it is now possible to probe entanglement and many-body correlations stored in a quantum state.

In this Perspective, we take a top-down route in terms of characterizing quantum information. First, we revisit the maximum speed at which quantum information propagates in a many-body system after a quench. How fast information propagates is intrinsically determined by the character of the inter-particle interactions and can be directly observable in the dynamics of two-body correlations \cite{Lieb1972,Bloch_LightCone_2012,Roos_EntangProp_2014,Monroe_Correlations_2014,Schmiedmayer_Lightcone_2013}. However, accessing higher order correlations is fundamental for gaining a full picture of  the dynamics of quantum information in many-body systems. We thus proceed to  discuss experimental developments on  how to  extract information of   many-body correlations. We do that by first  reviewing measurements based on quantum interference \cite{Schmiedmayer_ManyBody_2017}, and then discuss  new in-situ techniques accessible in state-of-the-art quantum gas microscopes \cite{Greiner_microscope_2009,Kuhr_microscope_2010, Herwig_microscopy_2008,Zwierlein_microscope_2015,Thywissen_microscope_2015,Bakr2017,Bloch_Review_2017,Yamamoto2016} and trapped ions\cite{Roos_IonsEntanglement_2018}. These latter techniques have allowed direct measurements of entanglement between different parts of a quantum system \cite{Greiner_Entanglement_2015,Roos_IonsEntanglement_2018}, to study  the role of entanglement in the emergence of statistical mechanics \cite{Kaufman_thermalization_2016,Roos_IonsEntanglement_2018}. Applying these techniques to systems that fail to thermalize --- known as many-body localized systems (MBL) ---  enabled new perspectives through studies of entanglement dynamics~\cite{Monroe_MBL_2016, Roos_IonsEntanglement_2018,Greiner_MBL_2018}, building on pioneering work with single-particle probes \cite{Bloch_MBL_2015,Choimany_2016,Bloch2d2017}. Finally, we discuss the intimately related topic of scrambling of quantum information, which is a concept first developed in attempts to understand the black hole information paradox, and fundamentally linked to how information dynamics leads to thermalization \cite{Shenker2014,Hayden2007,Sekino2008,Swingle2018, Qi2018}. We also discuss its characterization through out-of-time-order correlations. We conclude with a brief outlook, discussing some of the open questions which future work might address.

\section{General overview }

\subsection{Theoretical developments}

In quantum mechanics, interactions between particles can lead to the build-up of entanglement. A central question is how entanglement is distributed and how fast it spreads in non-equilibrium interacting quantum many-body systems. Answering  this question can impact quantum technologies, help designing optimal structures of quantum computer circuits and unveil the emergence of thermodynamics in isolated quantum systems\cite{Rigol_chaos_2016,Nandkishore_MBL_2015}.

The key to understanding the dynamics of quantum information is that it is stored in local degrees of freedom of the initial state of the system and can become scrambled across the global degrees of freedom of the system. This process is dubbed `information scrambling' \cite{Swingle2018}. Scrambling is generically accompanied by a build-up of many-body entanglement, causing the reduced density matrix of smaller subsystems to attain a steady state  that can be described by a statistical ensemble and thus to thermalize. The entanglement build-up, is therefore  the underlying reason why at the microscopic scale quantum mechanics can still lead to the emergence of behaviour consistent with statistical mechanics typically expected at the macroscopic scale.

One  associated question  then is how fast a system can thermalize. This is directly connected to the speed at which quantum information can propagate in a quantum system. For systems with only short-range interactions, Elliott Lieb and Derek William Robinson\cite{Lieb1972} derived a constant-velocity bound (now known as the Lieb-Robinson bound) that limits correlations to within a linear effective 'light cone',  similar to the linear  spreading of information in relativistic theories due to the finite speed of light. However, little is known about the propagation speed in systems with long-range interactions \cite{Hastings2006, Hauke2013,Eisert2013,FossFeig2015,Else2018}, partly because analytic solutions rarely exist and long-range interacting systems are very  hard to tackle with current numerical methods. New theoretical bounds have been derived from the study of black holes \cite{Maldacena2016}. They have led to the conjecture that fundamental bounds on quantum information spreading do exist for systems with generic interactions (for example interactions that decay as a power law with distance, such as dipolar or Coulomb interactions). However, it remains unclear if more familiar quantum many-body systems (such as those created in the cold atom laboratories ) saturate those bounds.

While most of the systems found in nature thermalize, there are  special types of systems which defy thermalization and can retain retrievable quantum correlations and avoid the spreading of quantum information to arbitrarily long times\cite{Nandkishore_MBL_2015,Altman2018}. Such MBL systems typically require the interplay of strong disorder and interactions\cite{Marko2008,Bardarson2012}.

\begin{figure}[htb!]
 \includegraphics[width=16cm]{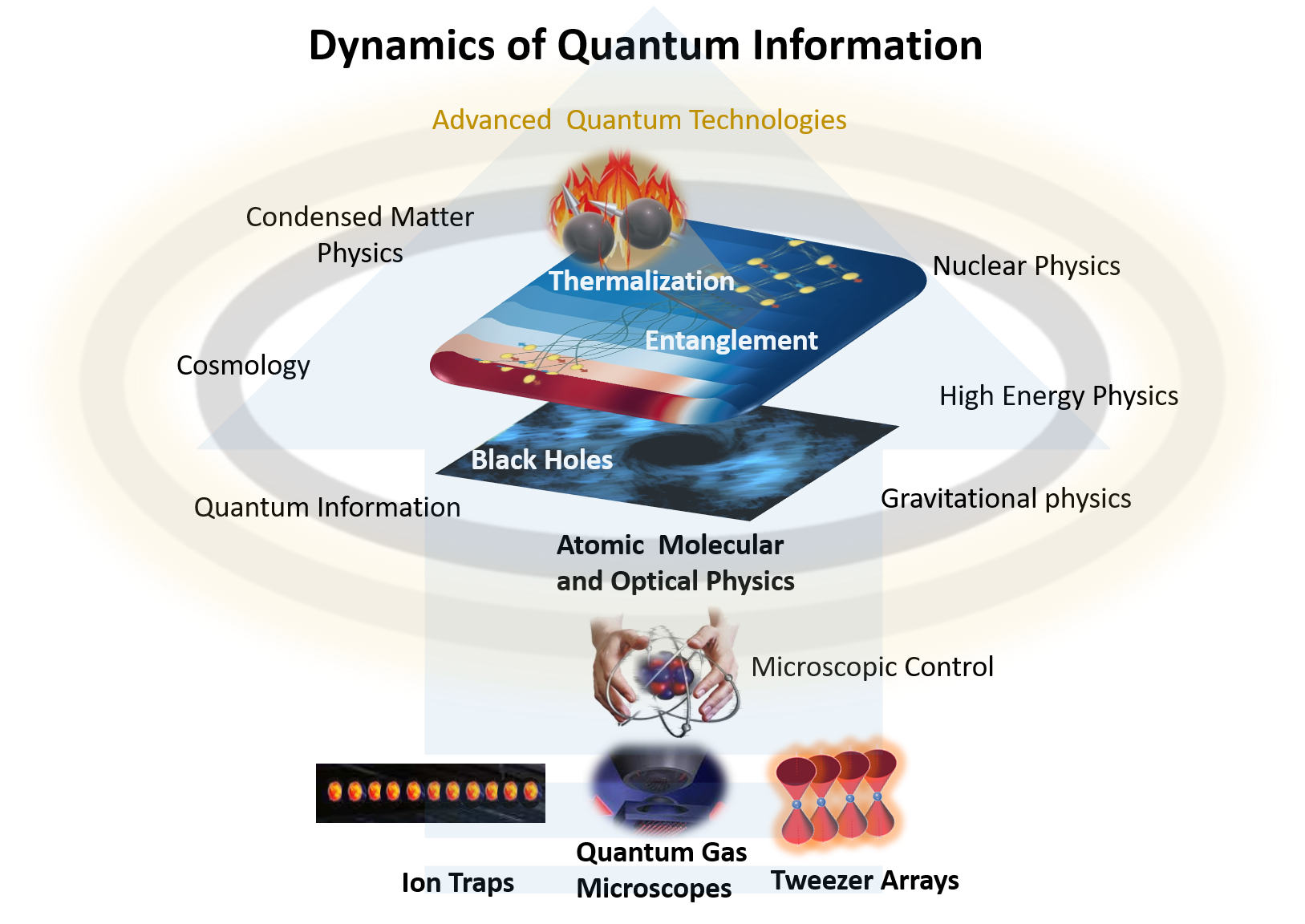}
 \caption{{\small Tools to control single atoms and ions enable us to probe, almost in real time, the dynamics of quantum information. Understanding the propagation of information and entanglement in complex systems is relevant for a broad range of disciplines with important fundamental and practical implications.  }}
 \label{fig:Overview} 
\end{figure}

\subsection{Experimental platforms}
The development of new experimental capabilities and methodologies has been  fundamental not only  to understanding bounds of propagation of  information in many-body systems  but also thermalization, scrambling and many-body localization. In particular,  synthetic quantum many-body  AMO systems -- including ultra-cold neutral atoms \cite{Bloch_Review_2017} and arrays of trapped ions\cite{Blatt_Review_2012} -- are playing an essential role in this effort. 
In these platforms experiments are starting to achieve  full microscopic control over single qubits encoded in  hyperfine states of neutral ground state atoms \cite{Greiner_microscope_2009,Kuhr_microscope_2010, Herwig_microscopy_2008,Zwierlein_microscope_2015,Thywissen_microscope_2015,Bakr2017,Bloch_Review_2017,Yamamoto2016,Regal2015}, trapped ions\cite{Roos_IonsEntanglement_2018,Zhang2017,Friis2018} and Rydberg states\cite{Gross_rydberg_2017,Lukin_51atomsimulator_2017,Browaeys_RydbergIsing_2016,Browaeys_IsingPRX_2018,Bakr2018}, combined with  tunable interactions between qubits.These interactions can be categorized in terms of their spatial range. Contact interactions, due to, for example, atomic collisions, are the dominant type of interactions in ultracold gases and are controllable by tuning  the scattering length $a_s$ via, for example, Feshbach resonances\cite{Julienne2010}. Various other interactions which decay as a power-law between particles separated by a distance $r$ are also accessible. For instance, van der Waals interactions $\propto 1/r^6$ can be realized in Rydberg atoms. Dipolar interactions $\propto 1/r^3$  \cite{Lahaye2009} , which additionally depend on the relative orientation of the interacting dipoles, are experienced by magnetic atoms, polar molecules and  Rydberg atoms. They are controllable by external electromagnetic fields. It is also possible to entirely engineer interactions  by coupling particle's internal  degrees of freedom (spin)  to shared bosonic modes 
\cite{Lev2018,Davis2019,Norcia2017}). For instance, when trapped ions are illuminated by laser beams a spin-dependent force can be created and used to virtually excite phonons in the ion crystal. The phonons in turn mediate spin-spin couplings that inherit the non-local structure of the collective modes. The diverse nature of these interactions enables the exploration of a broad parameter space and taking advantage of the complementary experimental capabilities offered by different platforms.

\section{Measuring correlation functions}

Lieb-Robinson bounds and the spatially-resolved propagation of quantum correlations were first observed and verified in a neutral atom quantum simulator of the 1D Bose-Hubbard model \cite{Bloch_LightCone_2012}.  A Bose gas of $^{87}$Rb trapped in an optical lattice was prepared in the Mott phase with one atom per site and  subsequently rapidly quenched to the superfluid phase, creating a non-equilibrium state which then was allowed to evolve. Non-trivial correlations built up in spatially distinct regions of the lattice, driven by the creation of local doublon/holon quasiparticle pairs. These quasiparticles propagated through the system with fixed opposite momenta and  at a constant characteristic velocity, consistent with and supporting the validity of the Lieb-Robinson theory for short-range interacting systems (see Fig.2a).

The investigation of similar bounds for long-range interacting systems, for which generic linear light-cone behaviour is not necessarily satisfied or expected \cite{Hastings2006, Hauke2013,Eisert2013,FossFeig2015,Else2018}, was pursued by experiments in chains of trapped ions \cite{Roos_EntangProp_2014,Monroe_Correlations_2014,Jurcevic2015}. Specifically,  for two spins, $i$ and $j$,   located at positions, ${r}_i$  and $r_j$  respectively, the investigated dynamics were  set by  1D XY and Ising spin models with spin couplings   decaying with interparticle distance, $|{r}_i-r_j|$, as a power-law with exponent $\alpha$: $J_{ij}\propto 1/|{r}_i-r_j|^\alpha$. Although the experiments reported in refs~\cite{Roos_EntangProp_2014,Monroe_Correlations_2014,Jurcevic2015} explored similar spin models, they differed  in the quench protocol used. refs~\cite{Roos_EntangProp_2014,Jurcevic2015} reported on the dynamics of a single excitation by flipping the spin of an individual ion in a polarized chain (Fig. 2b) and measured the propagation of information at later time via spatially-resolved two-particle correlations $C_{ij}(t) = \langle\hat \sigma^z_i \hat \sigma^z_j \rangle-\langle\hat \sigma^z_i \rangle\langle\hat \sigma^z_j \rangle$. Here $\hat \sigma^z_j $ are Pauli matrices acting on spin $j$. Although dynamics approximately consistent with a linear light-cone  was observed for $\alpha > 1.41$,  strong deviations from a linear wavefront were seen for smaller $\alpha$, signalling the breakdown of the simpler Lieb-Robinson type bound. Similar results were reported in ref.\cite{Monroe_Correlations_2014} but using a global quench where the dynamics of correlations must be thought of as the interference of many propagating quasi-particles, rather than a single excitation. These experimental results have later motivated theoretical progress in improving bounds for long-range interactions \cite{FossFeig2015}.

\begin{figure}[htb!]
 \includegraphics[width=12cm]{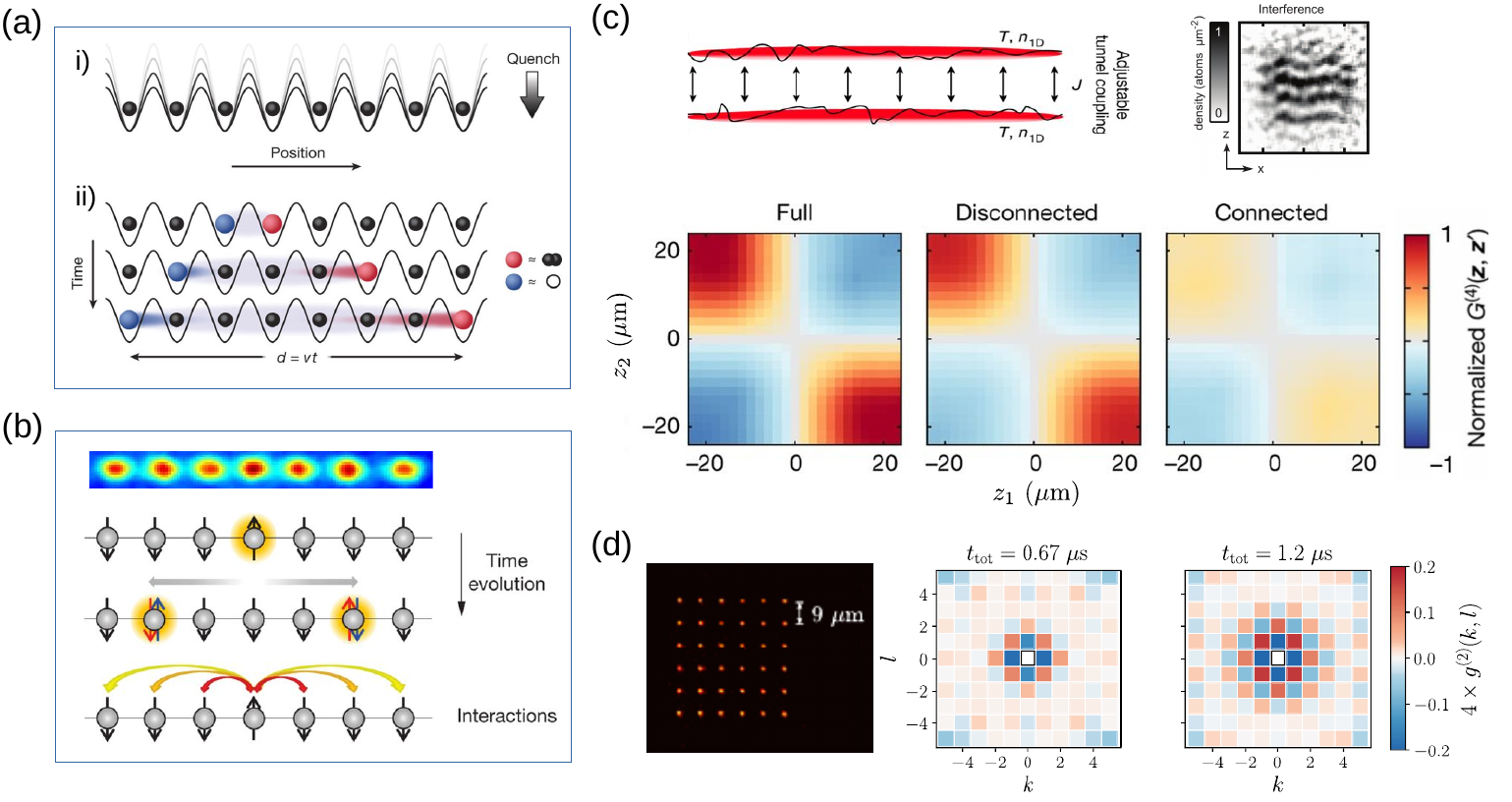}
 \caption{{\footnotesize Propagation and build-up of quantum correlations in non-equilibrium dynamics. (a) The propagation of correlations was investigated using neutral atoms in an optical lattice to emulate the Bose-Hubbard model. The atoms were initially prepared in a Mott state with commensurate filling. The trapping potential was quenched and led to non-equilibrium dynamics described by the creation of doublon (red circles) and holon (blue circles) quasi-particles, which created spatial correlations at a finite characteristic velocity. 
 (b) Similar correlation dynamics were explored in a 1D chain of trapped ions (false color image) \cite{Monroe_Correlations_2014, Roos_EntangProp_2014} which emulated a spin-model with variable long-range interactions. In particular, in  ref.~\cite{Roos_EntangProp_2014} non-trivial spin-spin correlations between spatially distinct regions of the ion chain were measured after the creation of a local excitation. 
 (c) Studying high-order correlations with tunnel coupled 1D Bose gases \cite{Schmiedmayer_ManyBody_2017}. Correlations between the phase profiles of two coupled 1D Bose gases were probed by matter-wave interference. An example $4$-th order two-point phase-correlation $G^{(4)}(z_1,z_2)$, where $z_1$ and $z_2$ are two different locations in the tube. The function $G^{(4)}(z_1,z_2)$ can be decomposed  into lower-order correlators, typically referred to as the  disconnected part, and non-factorizable terms, typically referred to as the connected contributions, which are a measure of the complexity of a quantum state. Atomic interactions and coupling of the 1D gases led to non-trivial buildup of correlations which cannot be factorized, manifested in a non-vanishing connected contribution. 
 (d) Correlation dynamics in Rydberg arrays. Optical tweezers were used to trap individual Rydberg atoms in a controllable array (left) \cite{Browaeys_IsingPRX_2018}. Effective spin-spin interactions were engineered via an atomic van der Waals interaction. Non-equilibrium dynamics were probed by spatially and time-resolved density-density correlations $g^{(2)}(k,l)=\frac{1}{N_{k,l}}\sum_{i,j} (\langle {\hat n}_i  {\hat n}_j \rangle  -\langle {\hat n}_i\rangle \langle  {\hat n}_j \rangle )$. Here ${\hat n}_i $ is a  projector on the Rydberg state for
atom $i$,  the sum runs over atom pairs $(i,j)$ whose separation
is ${\bf r}_i- {\bf  r}_j =(ka, la)$, $a$ is the lattice spacing and   $N_{k,l}$ is the number of such atom
pairs in the array (right). Panel (a) reproduced from ref.~\cite{Bloch_LightCone_2012}; panel (b) reproduced coutersy of the Innsbruck trapped ion group and refs.~\cite{Roos_EntangProp_2014}; panel (c) adapted from refs.~\cite{Schmiedmayer_ManyBody_2017,Schmiedmayer_GibssEnsemble_2015}; panel (d) adapted from Ref.~\cite{Browaeys_IsingPRX_2018}.}}
 \label{interfe}
\end{figure}

Propagation of correlations has also been studied in the context of the dynamics of a 1D Bose gas in ref.~\cite{Schmiedmayer_Lightcone_2013}. In this study, the focus was to demonstrate that relaxation of a many-body quantum system first develops  at local scales, due to the finite speed at which correlations can emerge between spatially separated points.

In parallel, the experiment also provided insights on how many-body quantum systems relax to steady states with local properties described within the framework of statistical mechanics. Leveraging the control and precision of their atom-chip setup, an initial equilibrium  quasi-condensate was split into two copies to form a highly non-equilibrium state. Using  matter-wave interference  (see Fig.2c), the group was able to probe the relaxation of the system towards a steady state by the measurement of two-point phase correlation functions. The steady state was observed to emerge at short length scales,  $z_c \sim 2ct$, (where $t$ is the time) bounded  by the finite characteristic speed $c$ at which correlations propagate through the system. This steady state is consistent with the predictions of a generalized Gibbs statistical ensemble (which accounted for the conserved quantities in the implemented model).

Subsequent advances in the atom-chip platform have led to the capability to characterize the experimental system in increasing detail, including up to 10-point correlation functions and full distribution functions. This ability was first used to further investigate the relaxation to a Gibbs ensemble \cite{Schmiedmayer_GibssEnsemble_2015}, before experiments studied quantum simulation of the sine-Gordon model with a pair of tunnel coupled 1D Bose gases \cite{Schmiedmayer_ManyBody_2017}.
The latter has opened a path to the characterization of the complexity of many-body states, by determining  the degree of non-factorizability of  high-order correlations into lower-order correlations (see Fig. 2c).

Arrays of optically-trapped Rydberg atoms enable the probing of the dynamics of correlations with single-atom resolution. By encoding an effective spin degree of freedom in the ground and excited Rydberg states, spin-spin interactions have been generated through strong van der Waals interactions between Rydberg atoms \cite{Browaeys_RydbergIsing_2016,Bakr2018,Lukin_51atomsimulator_2017,Browaeys_IsingPRX_2018,Keesling2018} and through optical dressing in a lattice \cite{Gross_rydberg_2017}, leading to the emulation of spin models in a new platform. In the former, the atoms are trapped in individual microtraps (tweezers) and excited to Rydberg states. The microtraps can be arranged such that the blockade radius $R_b$, that is, the distance over which inter-atomic interactions prevent the simultaneous excitation of two atoms, is comparable to the  separation between tweezers. This has allowed the study of rich  non-equilibrium dynamics following quenches \cite{Browaeys_RydbergIsing_2016,Bakr2018} and slow sweeps \cite{Lukin_51atomsimulator_2017,Keesling2018} in spin models with relatively short-range interactions. A quantum system of a nearest-neighbour Ising antiferromagnet was implemented in 1D and 2D neutral atom arrays \cite{Browaeys_RydbergIsing_2018,Bakr2018}, allowing the study of dynamics as experimental parameters were dynamically tuned  (see Fig.2d). Observations of non-trivial spatially-resolved spin-spin correlations exhibited a characteristic delay, which was used  as an experimental signature of the bounds on the propagation of correlations.

\section{Probing entanglement entropy}

So far, we have discussed how interactions can induce measurable correlations in a many-body system, and how these correlations can be subsequently measured in a variety of AMO platforms. However, these correlations alone do not always certify the presence of entanglement. Although in spin systems correlations in several bases of spin operators can quantify entanglement via full-state tomography, this approach scales poorly with size, limiting its applicability to many-body systems. Nevertheless, by capitalizing on recent advances in microscopic control of large AMO systems, a variety of new protocols with more flexible properties have allowed for direct measurement of entanglement dynamics.

These measurement protocols are built on the concept that entanglement fundamentally involves non-classical correlations between the different subsystems of a quantum many-body state. A subsystem might be delineated spatially, with respect to the momentum space,  or through other desirable partitions of the system. To connect with recent experimental studies, we focus on the case of entanglement between two spatially separated regions of a quantum state. Consider an initial non-entangled product state, $\vert \psi(0) \rangle = \vert \psi \rangle_A \otimes \vert \psi \rangle_B $ where $A$,$B$ refer to spatial subsystems of the full system, evolving under the interacting quantum many-body Hamiltonian. While the initial combined state of the quantum system may be written as a product of the state of pure subsystems, the Hamiltonian evolution may generate entanglement (see Fig. 3a). As a result, one will no longer be able to describe the state of each subsystem as a pure state, instead each is described by a mixed state due to the entropy induced by entanglement. The degree of mixedness can be quantified  in terms of the  von Neumann entanglement entropy $S_{vN} = \textrm{Tr}[\rho_A \log (\rho_A)]$ or the $n$-th order R\'enyi entropy $S_n  = \textrm{Tr}[\rho_A^n]$, where
$\rho_A$ is the reduced density matrix of the subsystem $A$. The growth of local entropy, $S_n$ or $S_{vN}$, in a closed, pure quantum system certifies the presence and quantifies the degree of entanglement present between the subsystems.

\begin{theo}[{Protocol for Measuring the R\'enyi Entropy}]{box:interference}
 \includegraphics[width=16cm]{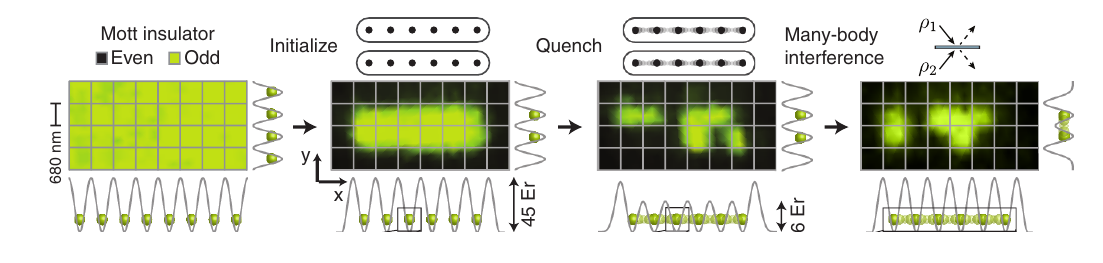}
\small{\noindent In a quantum gas microscope the R\'enyi entropy is measured as follows (a) After preparing a low-entropy Mott-insulator. (b) Two copies of a quantum state of interest are isolated. (c) The two states then each undergo an identical quench in the Hubbard parameters, by suddenly reducing the lattice depth. (d) The two copies are then interfered using a double-well beam-splitter interaction, and number-resolving measurements yield $\textrm{Tr}(\rho^2)$, with $\rho$ being the density matrix, for all subsystems and the full system simultaneously. In ref.~\cite{Kaufman_thermalization_2016}, this protocol was applied to a six site unity-filled Bose-Hubbard chain after a lattice quench from a Mott-insulator into the superfluid region of the ground-state phase diagram. This quench is meant to facilitate study of perturbing the system from equilibrium and observing its subsequent relaxation. Figure adapted with permission from ref.~\cite{Kaufman_thermalization_2016}.}
\end{theo}

The role entanglement entropy plays in the thermalization of closed quantum systems \cite{Alessio2015,cardyGS,cardyOneD} is particularly relevant. At first glance, quantum thermalization  may be a confusing concept: a unitarily evolving quantum system remains pure in time, which would seem to preclude the system's observables approaching those of an entropic thermal ensemble. However, as the isolated system evolves, the subsystems become more entropic due to becoming entangled with each other. In particular, when a quantum system thermalizes, the entanglement entropy scales extensively with the size of the subsystem, as expected from basic expectations of how thermal entropy should scale in statistical mechanics. In this scenario, the subsystems of a thermal ensemble and a highly entangled from the perspective of entanglement entropy pure state can become identical with respect to all local measurement observables. Hence, the entanglement entropy fulfills the role of the thermal entropy from statistical mechanics, so that sub-systems are faithfully described by maximum entropy ensembles. This crucial role of entanglement entropy has led to its use in differentiating phases of matter in non-equilibrium physics~\cite{Bardarson2012, Khemani2017}, such as thermalizing and many-body localizing phases. It also plays a role outside of quantum statistical mechanics in topological states of matter, where topological properties can induce entanglement entropy invariants from highly non-local correlations~\cite{Kitaev2006,Levin2006}.

To study these concepts experimentally, quantum gas microscopes have been used to generate pure quantum states that observably undergo this process of thermalization. In ref.~\cite{Kaufman_thermalization_2016}, using 6 bosons on 6 sites, the global purity and subsystem mixedness were measured using the proposed~\cite{Daley2012} and subsequently demonstrated technique of many-body interference~\cite{Greiner_Entanglement_2015, Kaufman_thermalization_2016}, which yields the second-order R\'enyi entropy $S_2 = \textrm{Tr}(\rho_A^2)$ (see Box 1 for details). It was observed that the entanglement entropy grew from a vanishing value consistent with the initial product state prepared, and subsequently saturated (with residual finite-size fluctuations) to a value near the expected thermal entropy of the sub-system (see Fig. 3b). At the same time, the full system entropy was observed to be static and near unity as a function of time, although more traditional local observables -- such as the on-site number distribution --  converged on the predictions stipulated by a thermal ensemble in the full-system eigenstates  (see Fig. 3b).
The scaling of the entanglement entropy at long times was contrasted with that of the ground-state for the same lattice parameters, illustrating the expected differences in the behavior of the entanglement entropy in the two regimes~\cite{cardyGS,cardyOneD,Alessio2015}.

An alternate protocol for measurement of the R\'enyi entropy involving randomized measurements combined with single particle resolution, was implemented for a trapped-ion quantum simulator~\cite{Roos_IonsEntanglement_2018}. The protocol allowed measurements of R\'enyi entropy for partitions up to  $10$ out of 20 ions. Similar to the quantum gas microscopes, the system can be initialized into a pure state with high fidelity. Furthermore, the sources of decoherence present  are well understood, allowing for local and global entropy to be distinguished and thus for entanglement dynamics to be properly characterized. This protocol requires only a single copy of a system and can be readily implemented in any experimental system where single particle addressability and detection are available.

 \begin{figure}[h!]
 \includegraphics[width=16cm]{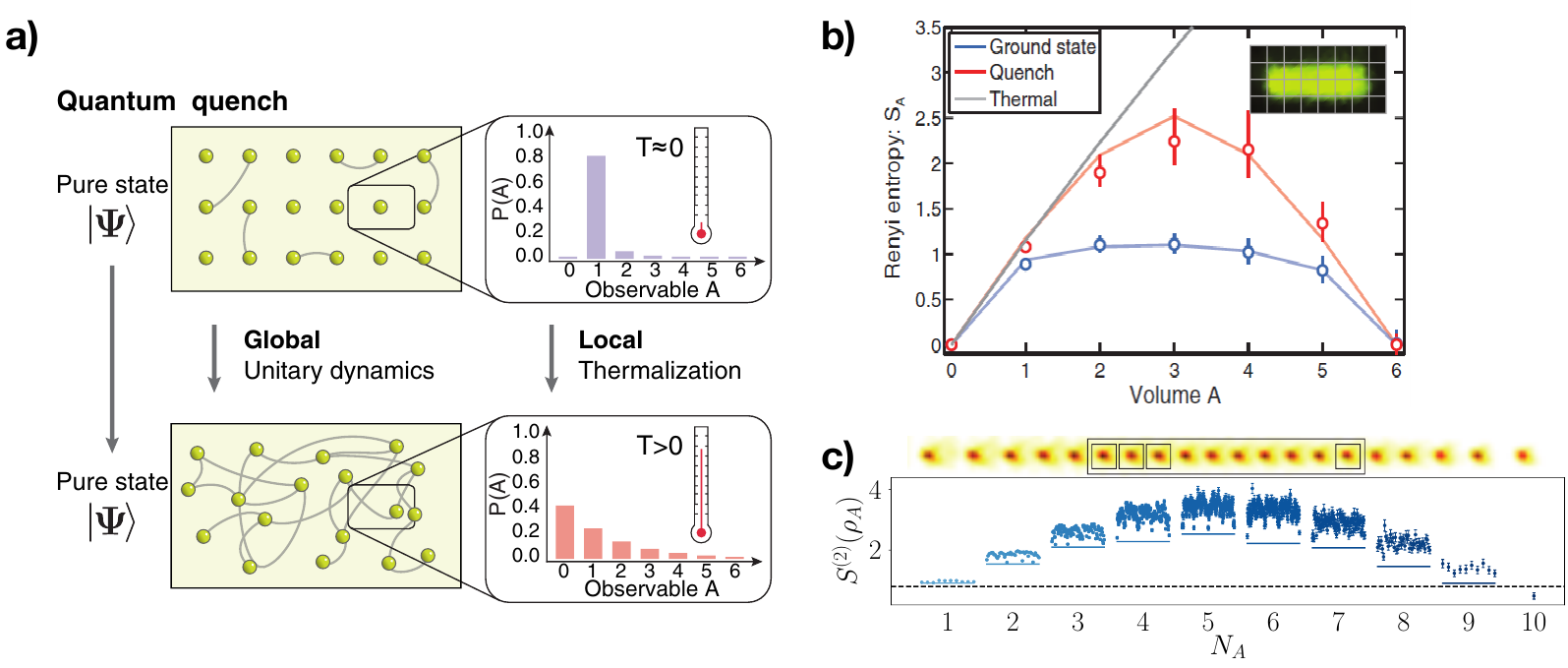}
 \caption{{\small Thermalization dynamics of an isolated quantum system accompanied by the build-up of entanglement entropy. (a) Schematic illustration of the thermalization process. In the top panel the system at temperature $T=0$ is described by a pure state $\vert \psi\rangle$ (ref. \cite{Kaufman_thermalization_2016}). If the state is close to a product state, then the entanglement (grey lines) between the different subsystems is negligible, so that each subsystem is also pure. The unitary evolution of the system post quench entangles all parts of the system, shown schematically in the bottom panel.   The bar graphs show the probability, $P(A)$, of an observable $A$
before and after perturbation of the system. Although
the full system remains in a pure and in
this sense zero-entropy state, the entropy of entanglement causes the subsystems to equilibrate, and local, thermal mixed states appear.
(b) and (c) The 'thermal' behaviour of the globally pure state after a quench  is manifest in the scaling of the entanglement entropy with subsystem size \cite{Roos_IonsEntanglement_2018}. In panel (b) the experimental data from a quantum gas microscope is shown; the red (blue) data illustrates the scaling of entanglement entropy, $S_A$, with subsystem size $A$ in the thermalized state (ground-state). Panel (c) illustrates measurements of the entanglement entropy, $S^2(\rho_A)$ vs subsystem size $N_A$, in a trapped-ion system using randomized measurement bases. In both cases, the scaling of the R\'enyi entropy for the thermalized state approaches that of a thermal ensemble. $\rho_A$ is the reduced density matrix of the subsystem $A$. Panels (a) and (b) adapted with ref. ~\cite{Kaufman_thermalization_2016}; panel (c) adapted from ref.~\cite{Roos_IonsEntanglement_2018}.}}
\label{fig:EntropyMeas}
\end{figure}

The MBL regime, where a sufficiently strong disorder prevents many-body interacting systems from thermalizing and suppresses the growth of entanglement entropy, has also been explored using quantum gas microscopes and ion traps. The primary signature of the MBL phase is the logarithmic growth of the sub-system entanglement entropy, $\mathcal S(t)\sim \log (t)$ in a model with nearest-neighbour couplings. However, a direct measurement of entanglement entropy is challenging, limiting experiments to systems of no more than $\sim 10-20$ particles, and to short times where the system remains coherent. Thus in the pioneering experiments done first in   1D quasi-random optical lattices \cite{Bloch_MBL_2015,Choimany_2016} with interacting fermions  and later in 
 a  two-dimensional array of interacting bosons  \cite{Bloch2d2017} a single body observable, imbalance $I(t)$, was used   to characterize the MBL phase. In the 1D case the  system was prepared in a charge density wave state with the odd lattice sites occupied, $N_{\rm odd}=N$, and $N_{\rm even}=0$. For this state, the dynamics of $I(t)=(N_{\rm odd}-N_{\rm even})/N$ was found to be  drastically different in the MBL phase, where the imbalance approached a constant non-zero value, versus the thermal phase, where it rapidly decayed to zero, signalling that all signatures of the initial order have vanished. Corresponding measurements of imbalance, adapted to the 2D case where theory is substantially more challenging, were performed in an interacting bosonic system via quantum gas microscopy~\cite{Bloch2d2017}, in which studies of two-point correlators were also possible. A similar approach was taken in the trapped-ion experiment in ref.~\cite{Monroe_MBL_2016}, where  a chain of 10 trapped ions was initialized in the Ne\'el state and subject to the transverse Ising Hamiltonian with power-law interactions, and with tunable quench disorder. The MBL phase was characterized by measuring the Hamming distance, which quantifies how many spin flips a state is different from the initial Ne\'el state. This observable is closely related to the quantum Fisher information, an entanglement witness, which has been shown to grow logarithmically in the MBL phase \cite{Bardarson2012}.

A clear experimental observation of the logarithmic growth of entanglement in the 
MBL phase was reported in ref.~\cite{Greiner_MBL_2018}, where the addition of site-resolved potential offset to the system in Ref.~\cite{Kaufman_thermalization_2016} allowed for the realization of the interacting Aubry-Andr\'e model. In this system the interaction strength, tunneling rate, and the disorder strength were varied to move between thermalizing and MBL phases. Two theoretically motivated quantities, namely the correlations between the spatial configuration of particles and the correlations between the number of particles in the two subsystems, were used as proxies for entanglement entropy. Furthermore, in ref.~\cite{Roos_IonsEntanglement_2018}, a direct measurement of the half-chain entanglement entropy dynamics showed marked difference between the thermalizing and the MBL phases. The large number of  theoretical investigations in the past few years  and subsequent experimental studies carried out highlight the continued critical role that entanglement plays in classifying many-body dynamics.

\section{Scrambling of quantum information}

Although time-ordered correlations display signatures of the apparent thermalization in closed quantum systems,  they do not capture the details of how information is 'scrambled', or  spread over the many-body degrees of freedom becoming inaccessible to  solely local probes. The simplest measures of scrambling are the expectations values associated with products of operators at different times, the lowest-order of which  has  been called out-of-time-order correlations (OTOCs). They are defined as   $C(t) = \langle \hat{W}^\dagger(t)\hat{V}^\dagger(0) \hat{W}(t) \hat{V}(0) \rangle$, where $\hat{V}(0)$ and $\hat{W}(0)$ are two commuting operators and  $\hat{W}(t)= e^{i \hat{H} t}  \hat{W}(0)  e^{-i \hat{H} t}$ the time evolved version of $\hat{W}(0)$ under the many-body Hamiltonian $\hat{H}$. The OTOC $C(t)$ can be interpreted as quantifying the non-commutativity of two initially commuting operators, a point which can be made explicit by noting the connection $\mathrm{Re}[C(t)] = 1 - \langle [\hat{W}(t),\hat{V}(0)]^{\dagger} [\hat{W}(t),\hat{V}(0)] \rangle/2$. OTOCs  are closely connected with the  spin-echo protocol introduced more than fifty years ago \cite{Hahn} and were first formally used in the context of superconductivity \cite{Larkin1969}. However, lately they have gained renewed attention given their key role in characterizing chaos, operator spreading and the scrambling of quantum information in many-body systems \cite{Fan2017,SwingleDisorderScrambling2017,Heyl_OTOCS_2018, Swingle_perspective_2018}.
Their study has opened a parallel new front for the understanding  of the dynamics of quantum information.

The generic tunability of AMO and nuclear magnetic resonance systems have enabled  pioneering experimental  measurements of OTOCs using spin-echo protocols. Here, by switching the sign of the Hamiltonian halfway through, the dynamics can effectively be reversed, essentially allowing the measurement of observables at different times, that is OTOCs. This has recently been achieved in macroscopic 2D arrays of trapped ions \cite{Rey_MQC_2017}, in a four-spin nuclear magnetic resonance system  (to see chaotic dynamics) \cite{Du_ScramblingNMR_2017},  in larger nuclear magnetic resonance chains (to see evidence of localization) \cite{Cappellaro2018},  in  momentum states of a Bose-Einstein condensate \cite{Bryce} and using a family of 3-qubit scrambling unitaries in an ion chain \cite{Monroe_Scrambling_2018}.

\begin{figure}[h!]
 \includegraphics[width=10cm]{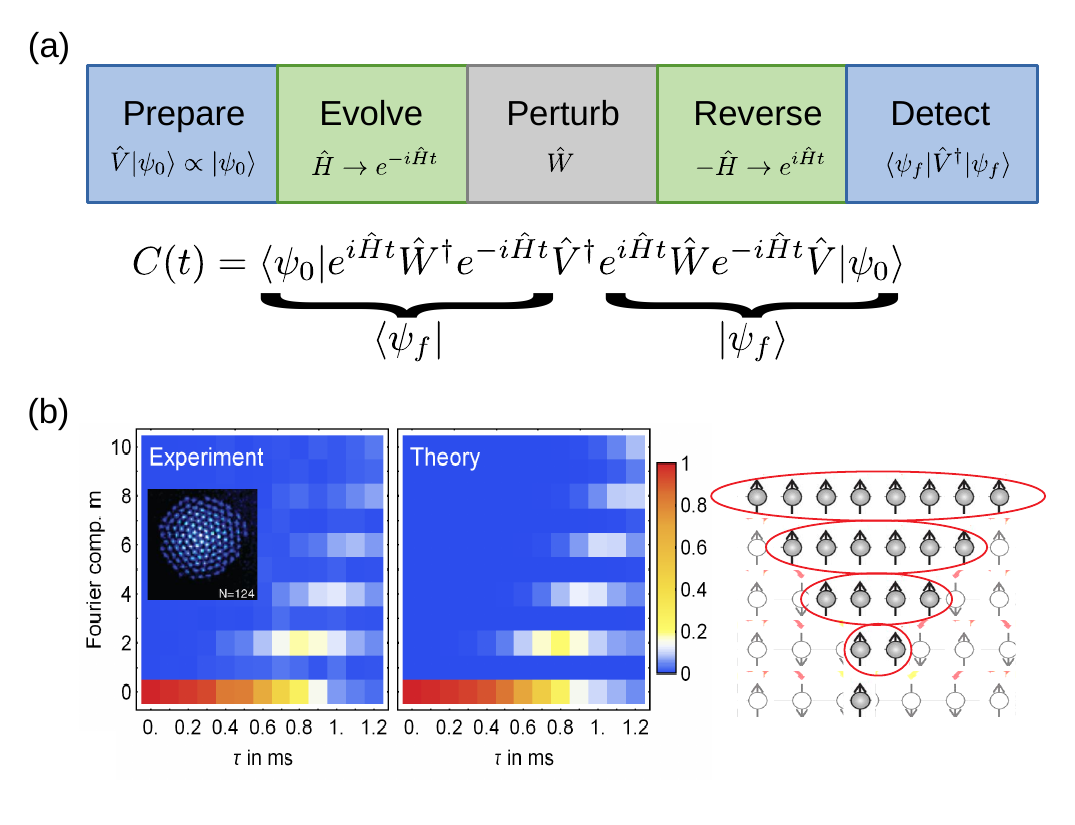}
 \caption{{\small Measurement and analysis of out-of-time-order correlations (OTOCs).
 (a) In atomic, molecular and optics platforms OTOCs have been measured with $\hat{V}$ chosen to be an operator  such that the initially prepared (pure) state is an eigenstate, $\hat{V}\vert \psi_0\rangle \propto \vert \psi_0 \rangle$. Control of the sign of the Hamiltonian $\hat{H} \to -\hat{H}$ allows the effective reversal of time and thus the OTOC $C(t)$ is reduced to measurement of $\hat{V}^{\dagger}$ for the time-evolved state $\vert \psi_f \rangle$. (b) In a 2D trapped ion array (inset) \cite{Rey_MQC_2017} the Fourier decomposition of an OTOC, $C_{\phi}(t) = \langle \hat{W}^{\dagger}_{\phi}(t) \hat{V}^{\dagger}(0) \hat{W}_{\phi}(t) \hat{V}(0)\rangle \equiv \sum_{m} A_m(t) e^{im\phi}$ was studied where $\hat{W}_{\phi}(0) = e^{-i\phi\hat{S}_x}$ was a global rotation and $\hat{V} = \hat{S}_x = \sum_j \hat{\sigma}^x_j/2$.Here, $\hat{\sigma}^x_j$  is the $x$ Pauli matrix acting on spin $i$. Observation of a non-zero Fourier amplitude $A_m \neq 0$ indicates the presence of $m$-body correlations between $m$ spins, as indicated by the grouped ions. Panel (b) adapted from Ref.~\cite{Rey_MQC_2017}.}}
 \label{scrambling}
\end{figure}

In these experiments the broad applicability and potential of OTOCs beyond just measuring scrambling was demonstrated. For example, single qubit control in ref.~\cite{Du_ScramblingNMR_2017} allowed the proof-of-principle reconstruction of entanglement entropy from measured OTOCs according to a proposal connecting these concepts \cite{Fan2017,Hosur2016}. Complementary to this, the trapped ion experiment demonstrated that the analysis of the Fourier decomposition of a set of OTOCs can be used to infer the buildup of $m$-body correlations between $m$ of the spins and to characterize  the growth of many-body coherences (see Fig. 4). A work in a  7-qubit ion trap  implemented a protocol to rigorously distinguish the effects of decoherence from scrambling \cite{Monroe_Scrambling_2018}.

These preliminary experiments have given just a small taste of the possible physics which AMO platforms could unveil through the study of OTOCs. Perhaps the most exciting prospect is to use engineered interactions in AMO systems to realize so-called fast scrambling models which might have connections to high-energy physics, such as the Sachdev-Ye-Kitaev model \cite{SYK}. Whilst links to classical chaos have indicated that for some quantum systems an OTOC can grow exponentially, $C(t) \sim e^{\lambda t}$, it has also been shown \cite{Maldacena2016} that generically this growth rate should be bounded in quantum systems by the temperature $\lambda \leq 2\pi T$. The study of OTOCs in high-energy physics has led to the intriguing insight that the scrambling rate in black holes saturates this bound, $\lambda = 2\pi T$, giving rise to the conjecture that any quantum system which similarly saturates the bound might be a holographic dual to a black hole \cite{Shenker2014,Hayden2007,Sekino2008,Shenker2015}. Exploring such connections using engineered analogue models in tabletop AMO experiments might lead to valuable insight in different areas.

\section{Outlook}

Despite the progress in understanding the role of quantum entanglement and correlations in the dynamics of many-body systems different directions remain to be explored. For example, measurements of entanglement between subsystems in quantum gas microscopes, microtraps and trapped ion systems have been constrained to only a handful of particles due to the increasing complexity of preparation, control, and detection with system size. The detailed characterization of the dynamics of quantum information via entanglement measurements in larger systems containing many tens to hundreds of particles, which are intractable to current theoretical methods,  remains an open challenge, and will require the design of new efficient and scalable alternative protocols for creating or measuring entanglement.

In parallel to the analog quantum simulators discussed here, the development of digital quantum simulators will enable the investigation of increasingly tunable and versatile physical systems \cite{Blatt_DigitalSimulation_2011}. These will open a path for the  study more complex problems that  saturate the bound of information scrambling or even violate it under some specific conditions \cite {Lucasgraph}. These include the realization of models relevant for high-energy physics \cite{Martinez2016}, investigation of fast scrambling \cite{Monroe_Scrambling_2018} and  future demonstrations of quantum supremacy via random unitary operations \cite{1203.5813,Keyserlingk2018,Khemani2018,Nahum2018}.

Finally, our discussion has been restricted to systems governed by time-independent  Hamiltonians. However, there is a variety of many-body phenomena which are being investigated in driven and dissipative regimes. These phenomena include Floquet dynamics\cite{Eckardt2017}, the emerging topic of time crystals \cite{Zhang:2017ci,Choi2017}, quantum synchronization \cite{Deutsch:2010ky,Solaro:2016iv,Piechon:2009cr,Norcia2017}, self-organization \cite{Baumann2010,Klinder2015,Leonard2017bis,Li:2017} and dynamical phase transitions \cite{Monroe_53Ion_2017,Jurcevic2017,Smale2018}. These  can provide a new understanding of the dynamics information in regimes beyond the unitary and time-independent Hamiltonians and thus be potentially    useful   for the design of optimal and robust protocols to store and transmit quantum information in many-body systems.

\vspace{5mm}
\noindent
{\bf Acknowledgments}\\  
We thank M. Norcia and A. Shankar for their careful reading of the manuscript and useful feedback.  This work is supported by the AFOSR grant FA9550-18-1-0319 and its MURI Initiative, by the DARPA and ARO grant W911NF-16-1-0576, the DARPA DRINQs program, the ARO single investigator award W911NF-19-1-0210,  the NSF PHY1820885, NSF JILA-PFC PHY-1734006 grants, and by NIST. 

\bibliographystyle{naturemag}

\begin{thebibliography}{10}
\expandafter\ifx\csname url\endcsname\relax
  \def\url#1{\texttt{#1}}\fi
\expandafter\ifx\csname urlprefix\endcsname\relax\def\urlprefix{URL }\fi
\providecommand{\bibinfo}[2]{#2}
\providecommand{\eprint}[2][]{\url{#2}}

\bibitem{Shenker2014}
\bibinfo{author}{Shenker, S.~H.} \& \bibinfo{author}{Stanford, D.}
\newblock \bibinfo{title}{Black holes and the butterfly effect}.
\newblock \emph{\bibinfo{journal}{Journal of High Energy Physics}}
  \textbf{\bibinfo{volume}{2014}}, \bibinfo{pages}{1--25}
  (\bibinfo{year}{2014}).

\bibitem{Hayden2007}
\bibinfo{author}{Hayden, P.} \& \bibinfo{author}{Preskill, J.}
\newblock \bibinfo{title}{Black holes as mirrors: quantum information in random
  subsystems}.
\newblock \emph{\bibinfo{journal}{Journal of High Energy Physics}}
  \textbf{\bibinfo{volume}{2007}}, \bibinfo{pages}{120} (\bibinfo{year}{2007}).

\bibitem{Sekino2008}
\bibinfo{author}{Sekino, Y.} \& \bibinfo{author}{Susskind, L.}
\newblock \bibinfo{title}{{Fast scramblers}}.
\newblock \emph{\bibinfo{journal}{J. High Energy Phys.}}
  \textbf{\bibinfo{volume}{2008}}, \bibinfo{pages}{065--065}
  (\bibinfo{year}{2008}).

\bibitem{Shenker2015}
\bibinfo{author}{Shenker, S.~H.} \& \bibinfo{author}{Stanford, D.}
\newblock \bibinfo{title}{Stringy effects in scrambling}.
\newblock \emph{\bibinfo{journal}{Journal of High Energy Physics}}
  \textbf{\bibinfo{volume}{2015}}, \bibinfo{pages}{1--34}
  (\bibinfo{year}{2015}).

\bibitem{Maldacena2003}
\bibinfo{author}{Maldacena, J.}
\newblock \bibinfo{title}{Eternal black holes in anti-de Sitter}.
\newblock \emph{\bibinfo{journal}{Journal of High Energy Physics}}
  \textbf{\bibinfo{volume}{2003}}, \bibinfo{pages}{021} (\bibinfo{year}{2003}).

\bibitem{Ryu2006}
\bibinfo{author}{Ryu, S.} \& \bibinfo{author}{Takayanagi, T.}
\newblock \bibinfo{title}{Holographic derivation of entanglement entropy from
  the anti--de Sitter space/conformal field theory correspondence}.
\newblock \emph{\bibinfo{journal}{Phys. Rev. Lett.}}
  \textbf{\bibinfo{volume}{96}}, \bibinfo{pages}{181602}
  (\bibinfo{year}{2006}).

\bibitem{Qi2018}
\bibinfo{author}{Qi, X.-L.}
\newblock \bibinfo{title}{Does gravity come from quantum information?}
\newblock \emph{\bibinfo{journal}{Nature Physics}}
  \textbf{\bibinfo{volume}{14}}, \bibinfo{pages}{984--987}
  (\bibinfo{year}{2018}).

\bibitem{Rigol_chaos_2016}
\bibinfo{author}{D'Alessio, L.}, \bibinfo{author}{Kafri, Y.},
  \bibinfo{author}{Polkovnikov, A.} \& \bibinfo{author}{Rigol, M.}
\newblock \bibinfo{title}{From quantum chaos and eigenstate thermalization to
  statistical mechanics and thermodynamics}.
\newblock \emph{\bibinfo{journal}{Advances in Physics}}
  \textbf{\bibinfo{volume}{65}}, \bibinfo{pages}{239--362}
  (\bibinfo{year}{2016}).

\bibitem{Nandkishore_MBL_2015}
\bibinfo{author}{Nandkishore, R.} \& \bibinfo{author}{Huse, D.~A.}
\newblock \bibinfo{title}{Many-body localization and thermalization in quantum
  statistical mechanics}.
\newblock \emph{\bibinfo{journal}{Annual Review of Condensed Matter Physics}}
  \textbf{\bibinfo{volume}{6}}, \bibinfo{pages}{15--38} (\bibinfo{year}{2015}).

\bibitem{Maldacena2016}
\bibinfo{author}{Maldacena, J.}, \bibinfo{author}{Shenker, S.~H.} \&
  \bibinfo{author}{Stanford, D.}
\newblock \bibinfo{title}{A bound on chaos}.
\newblock \emph{\bibinfo{journal}{Journal of High Energy Physics}}
  \textbf{\bibinfo{volume}{2016}}, \bibinfo{pages}{106} (\bibinfo{year}{2016}).

\bibitem{Hosur2016}
\bibinfo{author}{Hosur, P.}, \bibinfo{author}{Qi, X.-L.},
  \bibinfo{author}{Roberts, D.~A.} \& \bibinfo{author}{Yoshida, B.}
\newblock \bibinfo{title}{Chaos in quantum channels}.
\newblock \emph{\bibinfo{journal}{Journal of High Energy Physics}}
  \textbf{\bibinfo{volume}{2016}}, \bibinfo{pages}{1--49}
  (\bibinfo{year}{2016}).

\bibitem{Susskind2015}
\bibinfo{author}{Susskind, L.}
\newblock \bibinfo{title}{Computational complexity and black hole horizons}.
\newblock \emph{\bibinfo{journal}{Fortschritte der Physik}}
  \textbf{\bibinfo{volume}{64}}, \bibinfo{pages}{24--43}
  (\bibinfo{year}{2015}).

\bibitem{Lieb1972}
\bibinfo{author}{Lieb, E.~H.} \& \bibinfo{author}{Robinson, D.~W.}
\newblock \bibinfo{title}{The finite group velocity of quantum spin systems}.
\newblock \emph{\bibinfo{journal}{Communications in Mathematical Physics}}
  \textbf{\bibinfo{volume}{28}}, \bibinfo{pages}{251--257}
  (\bibinfo{year}{1972}).

\bibitem{Bloch_LightCone_2012}
\bibinfo{author}{Cheneau, M.} \emph{et~al.}
\newblock \bibinfo{title}{Light-cone-like spreading of correlations in a
  quantum many-body system}.
\newblock \emph{\bibinfo{journal}{Nature}} \textbf{\bibinfo{volume}{481}},
  \bibinfo{pages}{484--487} (\bibinfo{year}{2012}).

\bibitem{Roos_EntangProp_2014}
\bibinfo{author}{Jurcevic, P.} \emph{et~al.}
\newblock \bibinfo{title}{Quasiparticle engineering and entanglement
  propagation in a quantum many-body system}.
\newblock \emph{\bibinfo{journal}{Nature}} \textbf{\bibinfo{volume}{511}},
  \bibinfo{pages}{202--205} (\bibinfo{year}{2014}).

\bibitem{Monroe_Correlations_2014}
\bibinfo{author}{Richerme, P.} \emph{et~al.}
\newblock \bibinfo{title}{Non-local propagation of correlations in quantum
  systems with long-range interactions}.
\newblock \emph{\bibinfo{journal}{Nature}} \textbf{\bibinfo{volume}{511}},
  \bibinfo{pages}{198--201} (\bibinfo{year}{2014}).

\bibitem{Schmiedmayer_Lightcone_2013}
\bibinfo{author}{Langen, T.}, \bibinfo{author}{Geiger, R.},
  \bibinfo{author}{Kuhnert, M.}, \bibinfo{author}{Rauer, B.} \&
  \bibinfo{author}{Schmiedmayer, J.}
\newblock \bibinfo{title}{Local emergence of thermal correlations in an
  isolated quantum many-body system}.
\newblock \emph{\bibinfo{journal}{Nature Physics}}
  \textbf{\bibinfo{volume}{9}}, \bibinfo{pages}{640--643} (\bibinfo{year}{2013}).

\bibitem{Schmiedmayer_ManyBody_2017}
\bibinfo{author}{Schweigler, T.} \emph{et~al.}
\newblock \bibinfo{title}{Experimental characterization of a quantum many-body
  system via higher-order correlations}.
\newblock \emph{\bibinfo{journal}{Nature}} \textbf{\bibinfo{volume}{545}},
  \bibinfo{pages}{323--326} (\bibinfo{year}{2017}).

\bibitem{Greiner_microscope_2009}
\bibinfo{author}{Bakr, W.~S.}, \bibinfo{author}{Gillen, J.~I.},
  \bibinfo{author}{Peng, A.}, \bibinfo{author}{F{\"o}lling, S.} \&
  \bibinfo{author}{Greiner, M.}
\newblock \bibinfo{title}{A quantum gas microscope for detecting single atoms
  in a Hubbard-regime optical lattice}.
\newblock \emph{\bibinfo{journal}{Nature}} \textbf{\bibinfo{volume}{462}},
  \bibinfo{pages}{74--77} (\bibinfo{year}{2009}).

\bibitem{Kuhr_microscope_2010}
\bibinfo{author}{Sherson, J.~F.} \emph{et~al.}
\newblock \bibinfo{title}{Single-atom-resolved fluorescence imaging of an
  atomic Mott insulator}.
\newblock \emph{\bibinfo{journal}{Nature}} \textbf{\bibinfo{volume}{467}},
  \bibinfo{pages}{68-72} (\bibinfo{year}{2010}).

\bibitem{Herwig_microscopy_2008}
\bibinfo{author}{Gericke, T.}, \bibinfo{author}{W{\"u}rtz, P.},
  \bibinfo{author}{Reitz, D.}, \bibinfo{author}{Langen, T.} \&
  \bibinfo{author}{Ott, H.}
\newblock \bibinfo{title}{High-resolution scanning electron microscopy of an
  ultracold quantum gas}.
\newblock \emph{\bibinfo{journal}{Nature Physics}}
  \textbf{\bibinfo{volume}{4}}, \bibinfo{pages}{949--953} (\bibinfo{year}{2008}).

\bibitem{Zwierlein_microscope_2015}
\bibinfo{author}{Cheuk, L.~W.} \emph{et~al.}
\newblock \bibinfo{title}{Quantum-gas microscope for fermionic atoms}.
\newblock \emph{\bibinfo{journal}{Phys. Rev. Lett.}}
  \textbf{\bibinfo{volume}{114}}, \bibinfo{pages}{193001}
  (\bibinfo{year}{2015}).

\bibitem{Thywissen_microscope_2015}
\bibinfo{author}{Edge, G. J.~A.} \emph{et~al.}
\newblock \bibinfo{title}{Imaging and addressing of individual fermionic atoms
  in an optical lattice}.
\newblock \emph{\bibinfo{journal}{Phys. Rev. A}} \textbf{\bibinfo{volume}{92}},
  \bibinfo{pages}{063406} (\bibinfo{year}{2015}).

\bibitem{Bakr2017}
\bibinfo{author}{Mitra, D.} \emph{et~al.}
\newblock \bibinfo{title}{Quantum gas microscopy of an attractive
  Fermi-Hubbard system}.
\newblock \emph{\bibinfo{journal}{Nature Physics}}
  \textbf{\bibinfo{volume}{14}}, \bibinfo{pages}{173--177} (\bibinfo{year}{2017}).

\bibitem{Bloch_Review_2017}
\bibinfo{author}{Gross, C.} \& \bibinfo{author}{Bloch, I.}
\newblock \bibinfo{title}{Quantum simulations with ultracold atoms in optical
  lattices}.
\newblock \emph{\bibinfo{journal}{Science}} \textbf{\bibinfo{volume}{357}},
  \bibinfo{pages}{995--1001} (\bibinfo{year}{2017}).

\bibitem{Yamamoto2016}
\bibinfo{author}{Yamamoto, R.}, \bibinfo{author}{Kobayashi, J.},
  \bibinfo{author}{Kuno, T.}, \bibinfo{author}{Kato, K.} \&
  \bibinfo{author}{Takahashi, Y.}
\newblock \bibinfo{title}{An ytterbium quantum gas microscope with narrow-line
  laser cooling}.
\newblock \emph{\bibinfo{journal}{New Journal of Physics}}
  \textbf{\bibinfo{volume}{18}}, \bibinfo{pages}{023016}
  (\bibinfo{year}{2016}).

\bibitem{Greiner_Entanglement_2015}
\bibinfo{author}{Islam, R.} \emph{et~al.}
\newblock \bibinfo{title}{Measuring entanglement entropy in a quantum many-body
  system}.
\newblock \emph{\bibinfo{journal}{Nature}} \textbf{\bibinfo{volume}{528}},
  \bibinfo{pages}{77--83} (\bibinfo{year}{2015}).

\bibitem{Roos_IonsEntanglement_2018}
\bibinfo{author}{Brydges, T.} \emph{et~al.}
\newblock \bibinfo{title}{Probing Renyi entanglement entropy via randomized
  measurements}.
\newblock \emph{\bibinfo{journal}{Science}} \textbf{\bibinfo{volume}{364}},
  \bibinfo{pages}{260--263} (\bibinfo{year}{2019}).

\bibitem{Kaufman_thermalization_2016}
\bibinfo{author}{Kaufman, A.~M.} \emph{et~al.}
\newblock \bibinfo{title}{Quantum thermalization through entanglement in an
  isolated many-body system}.
\newblock \emph{\bibinfo{journal}{Science}} \textbf{\bibinfo{volume}{353}},
  \bibinfo{pages}{794--800} (\bibinfo{year}{2016}).
\newblock \eprint{http://science.sciencemag.org/content/353/6301/794.full.pdf}.

\bibitem{Bloch_MBL_2015}
\bibinfo{author}{Schreiber, M.} \emph{et~al.}
\newblock \bibinfo{title}{Observation of many-body localization of interacting
  fermions in a quasirandom optical lattice}.
\newblock \emph{\bibinfo{journal}{Science}} \textbf{\bibinfo{volume}{349}},
  \bibinfo{pages}{842--845} (\bibinfo{year}{2015}).

\bibitem{Choimany_2016}
\bibinfo{author}{Choi, J.-y.} \emph{et~al.}
\newblock \bibinfo{title}{Exploring the many-body localization transition in
  two dimensions}.
\newblock \emph{\bibinfo{journal}{Science}} \textbf{\bibinfo{volume}{352}},
  \bibinfo{pages}{1547--1552} (\bibinfo{year}{2016}).


\bibitem{Monroe_MBL_2016}
\bibinfo{author}{Smith, J.} \emph{et~al.}
\newblock \bibinfo{title}{Many-body localization in a quantum simulator with
  programmable random disorder}.
\newblock \emph{\bibinfo{journal}{Nature Physics}}
  \textbf{\bibinfo{volume}{12}}, \bibinfo{pages}{907--911} (\bibinfo{year}{2016}).

\bibitem{Greiner_MBL_2018}
\bibinfo{author}{Lukin, A.} \emph{et~al.}
\newblock \bibinfo{title}{Probing entanglement in a many-body-localized system}
  (\bibinfo{year}{2018}).
\newblock \eprint{arXiv:1805.09819}.

\bibitem{Bloch2d2017}
\bibinfo{author}{L\"uschen, H.~P.} \emph{et~al.}
\newblock \bibinfo{title}{Signatures of many-body localization in a controlled
  open quantum system}.
\newblock \emph{\bibinfo{journal}{Phys. Rev. X}} \textbf{\bibinfo{volume}{7}},
  \bibinfo{pages}{011034} (\bibinfo{year}{2017}).

\bibitem{Swingle2018}
\bibinfo{author}{Swingle, B.}
\newblock \bibinfo{title}{Unscrambling the physics of out-of-time-order
  correlators}.
\newblock \emph{\bibinfo{journal}{Nature Physics}}
  \textbf{\bibinfo{volume}{14}}, \bibinfo{pages}{988--990}
  (\bibinfo{year}{2018}).

\bibitem{Hastings2006}
\bibinfo{author}{Hastings, M.~B.} \& \bibinfo{author}{Koma, T.}
\newblock \bibinfo{title}{Spectral gap and exponential decay of correlations}.
\newblock \emph{\bibinfo{journal}{Communications in Mathematical Physics}}
  \textbf{\bibinfo{volume}{265}}, \bibinfo{pages}{781--804}
  (\bibinfo{year}{2006}).

\bibitem{Hauke2013}
\bibinfo{author}{Hauke, P.} \& \bibinfo{author}{Tagliacozzo, L.}
\newblock \bibinfo{title}{Spread of correlations in long-range interacting
  quantum systems}.
\newblock \emph{\bibinfo{journal}{Phys. Rev. Lett.}}
  \textbf{\bibinfo{volume}{111}}, \bibinfo{pages}{207202}
  (\bibinfo{year}{2013}).

\bibitem{Eisert2013}
\bibinfo{author}{Eisert, J.}, \bibinfo{author}{van~den Worm, M.},
  \bibinfo{author}{Manmana, S.~R.} \& \bibinfo{author}{Kastner, M.}
\newblock \bibinfo{title}{Breakdown of quasilocality in long-range quantum
  lattice models}.
\newblock \emph{\bibinfo{journal}{Phys. Rev. Lett.}}
  \textbf{\bibinfo{volume}{111}}, \bibinfo{pages}{260401}
  (\bibinfo{year}{2013}).

\bibitem{FossFeig2015}
\bibinfo{author}{Foss-Feig, M.}, \bibinfo{author}{Gong, Z.-X.},
  \bibinfo{author}{Clark, C.~W.} \& \bibinfo{author}{Gorshkov, A.~V.}
\newblock \bibinfo{title}{Nearly linear light cones in long-range interacting
  quantum systems}.
\newblock \emph{\bibinfo{journal}{Phys. Rev. Lett.}}
  \textbf{\bibinfo{volume}{114}}, \bibinfo{pages}{157201}
  (\bibinfo{year}{2015}).

\bibitem{Else2018}
\bibinfo{author}{Else, D.~V.}, \bibinfo{author}{Machado, F.},
  \bibinfo{author}{Nayak, C.} \& \bibinfo{author}{Yao, Y.}
\newblock \bibinfo{title}{An improved Lieb-Robinson bound for many-body
  hamiltonians with power-law interactions}.
\newblock \emph{\bibinfo{journal}{Arxiv}} \textbf{\bibinfo{volume}{1809.06369}}
  (\bibinfo{year}{2018}).

\bibitem{Altman2018}
\bibinfo{author}{Altman, E.}
\newblock \bibinfo{title}{Many-body localization and quantum thermalization}.
\newblock \emph{\bibinfo{journal}{Nature Physics}}
  \textbf{\bibinfo{volume}{14}}, \bibinfo{pages}{979--983}
  (\bibinfo{year}{2018}).

\bibitem{Marko2008}
\bibinfo{author}{\ifmmode \check{Z}\else
  \v{Z}\fi{}nidari\ifmmode~\check{c}\else \v{c}\fi{}, M.},
  \bibinfo{author}{Prosen, T. c.~v.} \&
  \bibinfo{author}{Prelov\ifmmode~\check{s}\else \v{s}\fi{}ek, P.}
\newblock \bibinfo{title}{Many-body localization in the Heisenberg $xxz$ magnet
  in a random field}.
\newblock \emph{\bibinfo{journal}{Phys. Rev. B}} \textbf{\bibinfo{volume}{77}},
  \bibinfo{pages}{064426} (\bibinfo{year}{2008}).

\bibitem{Bardarson2012}
\bibinfo{author}{Bardarson, J.~H.}, \bibinfo{author}{Pollmann, F.} \&
  \bibinfo{author}{Moore, J.~E.}
\newblock \bibinfo{title}{Unbounded growth of entanglement in models of
  many-body localization}.
\newblock \emph{\bibinfo{journal}{Phys. Rev. Lett.}}
  \textbf{\bibinfo{volume}{109}}, \bibinfo{pages}{017202}
  (\bibinfo{year}{2012}).

\bibitem{Blatt_Review_2012}
\bibinfo{author}{Blatt, R.} \& \bibinfo{author}{Roos, C.~F.}
\newblock \bibinfo{title}{Quantum simulations with trapped ions}.
\newblock \emph{\bibinfo{journal}{Nature Physics}}
  \textbf{\bibinfo{volume}{8}}, \bibinfo{pages}{277--284} (\bibinfo{year}{2012}).

\bibitem{Regal2015}
\bibinfo{author}{Kaufman, A.~M.} \emph{et~al.}
\newblock \bibinfo{title}{Entangling two transportable neutral atoms via local
  spin exchange}.
\newblock \emph{\bibinfo{journal}{Nature}} \textbf{\bibinfo{volume}{527}},
  \bibinfo{pages}{208--211} (\bibinfo{year}{2015}).

\bibitem{Zhang2017}
\bibinfo{author}{Zhang, J.} \emph{et~al.}
\newblock \bibinfo{title}{Observation of a many-body dynamical phase transition
  with a 53-qubit quantum simulator}.
\newblock \emph{\bibinfo{journal}{Nature}} \textbf{\bibinfo{volume}{551}},
  \bibinfo{pages}{601--604} (\bibinfo{year}{2017}).

\bibitem{Friis2018}
\bibinfo{author}{Friis, N.} \emph{et~al.}
\newblock \bibinfo{title}{Observation of entangled states of a fully controlled
  20-qubit system}.
\newblock \emph{\bibinfo{journal}{Phys. Rev. X}} \textbf{\bibinfo{volume}{8}},
  \bibinfo{pages}{021012} (\bibinfo{year}{2018}).

\bibitem{Gross_rydberg_2017}
\bibinfo{author}{Zeiher, J.} \emph{et~al.}
\newblock \bibinfo{title}{Coherent many-body spin dynamics in a long-range
  interacting Ising chain}.
\newblock \emph{\bibinfo{journal}{Phys. Rev. X}} \textbf{\bibinfo{volume}{7}},
  \bibinfo{pages}{041063} (\bibinfo{year}{2017}).

\bibitem{Lukin_51atomsimulator_2017}
\bibinfo{author}{Bernien, H.} \emph{et~al.}
\newblock \bibinfo{title}{Probing many-body dynamics on a 51-atom quantum
  simulator}.
\newblock \emph{\bibinfo{journal}{Nature}} \textbf{\bibinfo{volume}{551}}, \bibinfo{volume}{579--584}  (\bibinfo{year}{2017}).

\bibitem{Browaeys_RydbergIsing_2016}
\bibinfo{author}{Labuhn, H.} \emph{et~al.}
\newblock \bibinfo{title}{Tunable two-dimensional arrays of single Rydberg
  atoms for realizing quantum sing models}.
\newblock \emph{\bibinfo{journal}{Nature}} \textbf{\bibinfo{volume}{534}},
  \bibinfo{pages}{667--670} (\bibinfo{year}{2016}).

\bibitem{Browaeys_IsingPRX_2018}
\bibinfo{author}{Lienhard, V.} \emph{et~al.}
\newblock \bibinfo{title}{Observing the space- and time-dependent growth of
  correlations in dynamically tuned synthetic Ising models with
  antiferromagnetic interactions}.
\newblock \emph{\bibinfo{journal}{Phys. Rev. X}} \textbf{\bibinfo{volume}{8}},
  \bibinfo{pages}{021070} (\bibinfo{year}{2018}).

\bibitem{Bakr2018}
\bibinfo{author}{Guardado-Sanchez, E.} \emph{et~al.}
\newblock \bibinfo{title}{Probing the quench dynamics of antiferromagnetic
  correlations in a 2d quantum Ising spin system}.
\newblock \emph{\bibinfo{journal}{Phys. Rev. X}} \textbf{\bibinfo{volume}{8}},
  \bibinfo{pages}{021069} (\bibinfo{year}{2018}).

\bibitem{Julienne2010}
\bibinfo{author}{Chin, C.}, \bibinfo{author}{Grimm, R.},
  \bibinfo{author}{Julienne, P.} \& \bibinfo{author}{Tiesinga, E.}
\newblock \bibinfo{title}{Feshbach resonances in ultracold gases}.
\newblock \emph{\bibinfo{journal}{Rev. Mod. Phys.}}
  \textbf{\bibinfo{volume}{82}}, \bibinfo{pages}{1225--1286}
  (\bibinfo{year}{2010}).

\bibitem{Lahaye2009}
\bibinfo{author}{Lahaye, T.}, \bibinfo{author}{Menotti, C.},
  \bibinfo{author}{Santos, L.}, \bibinfo{author}{Lewenstein, M.} \&
  \bibinfo{author}{Pfau, T.}
\newblock \bibinfo{title}{The physics of dipolar bosonic quantum gases}.
\newblock \emph{\bibinfo{journal}{Reports on Progress in Physics}}
  \textbf{\bibinfo{volume}{72}}, \bibinfo{pages}{126401}
  (\bibinfo{year}{2009}).

\bibitem{Lev2018}
\bibinfo{author}{Vaidya, V.~D.} \emph{et~al.}
\newblock \bibinfo{title}{Tunable-range, photon-mediated atomic interactions in
  multimode cavity QED}.
\newblock \emph{\bibinfo{journal}{Phys. Rev. X}} \textbf{\bibinfo{volume}{8}},
  \bibinfo{pages}{011002} (\bibinfo{year}{2018}).

\bibitem{Davis2019}
\bibinfo{author}{Davis, E.~J.}, \bibinfo{author}{Bentsen, G.},
  \bibinfo{author}{Homeier, L.}, \bibinfo{author}{Li, T.} \&
  \bibinfo{author}{Schleier-Smith, M.~H.}
\newblock \bibinfo{title}{Photon-mediated spin-exchange dynamics of spin-1
  atoms}.
\newblock \emph{\bibinfo{journal}{Phys. Rev. Lett.}}
  \textbf{\bibinfo{volume}{122}}, \bibinfo{pages}{010405}
  (\bibinfo{year}{2019}).

\bibitem{Norcia2017}
\bibinfo{author}{Norcia, M.~A.} \emph{et~al.}
\newblock \bibinfo{title}{Cavity mediated collective spin exchange interactions
  in a strontium superradiant laser}.
\newblock \emph{\bibinfo{journal}{Science}} \textbf{\bibinfo{volume}{361}},
  \bibinfo{pages}{259--262} (\bibinfo{year}{2017}).

\bibitem{Jurcevic2015}
\bibinfo{author}{Jurcevic, P.} \emph{et~al.}
\newblock \bibinfo{title}{Spectroscopy of interacting quasiparticles in trapped
  ions}.
\newblock \emph{\bibinfo{journal}{Phys. Rev. Lett.}}
  \textbf{\bibinfo{volume}{115}}, \bibinfo{pages}{100501}
  (\bibinfo{year}{2015}).

\bibitem{Schmiedmayer_GibssEnsemble_2015}
\bibinfo{author}{Langen, T.} \emph{et~al.}
\newblock \bibinfo{title}{Experimental observation of a generalized Gibbs
  ensemble}.
\newblock \emph{\bibinfo{journal}{Science}} \textbf{\bibinfo{volume}{348}},
  \bibinfo{pages}{207--211} (\bibinfo{year}{2015}).

\bibitem{Keesling2018}
\bibinfo{author}{Keesling, A.} \emph{et~al.}
\newblock \bibinfo{title}{Probing quantum critical dynamics on a programmable
 Rrydberg simulator}.
\newblock \emph{\bibinfo{journal}{arXiv:1809.05540}}  (\bibinfo{year}{2018}).

\bibitem{Browaeys_RydbergIsing_2018}
\bibinfo{author}{Lienhard, V.} \emph{et~al.}
\newblock \bibinfo{title}{Observing the space- and time-dependent growth of
  correlations in dynamically tuned synthetic Ising models with
  antiferromagnetic interactions}.
\newblock \emph{\bibinfo{journal}{Phys. Rev. X}} \textbf{\bibinfo{volume}{8}},
  \bibinfo{pages}{021070} (\bibinfo{year}{2018}).

\bibitem{InnsbruckWebsite}
\bibinfo{title}{Ion traps for quantum information}.
\newblock
  \bibinfo{howpublished}{\url{https://quantumoptics.at/en/news/72-scalable-multiparticle-entanglement-of-trapped-ions.html}}.
\newblock \bibinfo{note}{2005 (accessed Novemeber 2018)}.

\bibitem{Alessio2015}
\bibinfo{author}{D'Alessio, L.}, \bibinfo{author}{Kafri, Y.},
  \bibinfo{author}{Polkovnikov, A.} \& \bibinfo{author}{Rigol, M.}
\newblock \bibinfo{title}{From quantum chaos and eigenstate thermalization to
  statistical mechanics and thermodynamics}.
\newblock \emph{\bibinfo{journal}{Advances in Physics}}
  \textbf{\bibinfo{volume}{65}}, \bibinfo{pages}{239--362}
  (\bibinfo{year}{2016}).


\bibitem{cardyGS}
\bibinfo{author}{Calabrese, P.} \& \bibinfo{author}{Cardy, J.}
\newblock \bibinfo{title}{Entanglement entropy and quantum field theory}.
\newblock \emph{\bibinfo{journal}{Journal of Statistical Mechanics: Theory and
  Experiment}} \textbf{\bibinfo{volume}{2004}}, \bibinfo{pages}{P06002}
  (\bibinfo{year}{2004}).

\bibitem{cardyOneD}
\bibinfo{author}{Calabrese, P.} \& \bibinfo{author}{Cardy, J.}
\newblock \bibinfo{title}{Evolution of entanglement entropy in one-dimensional
  systems}.
\newblock \emph{\bibinfo{journal}{Journal of Statistical Mechanics: Theory and
  Experiment}} \textbf{\bibinfo{volume}{2005}}, \bibinfo{pages}{P04010}
  (\bibinfo{year}{2005}).

\bibitem{Khemani2017}
\bibinfo{author}{Khemani, V.}, \bibinfo{author}{Lim, S.~P.},
  \bibinfo{author}{Sheng, D.~N.} \& \bibinfo{author}{Huse, D.~A.}
\newblock \bibinfo{title}{Critical properties of the many-body localization
  transition}.
\newblock \emph{\bibinfo{journal}{Phys. Rev. X}} \textbf{\bibinfo{volume}{7}},
  \bibinfo{pages}{021013} (\bibinfo{year}{2017}).

\bibitem{Kitaev2006}
\bibinfo{author}{Kitaev, A.} \& \bibinfo{author}{Preskill, J.}
\newblock \bibinfo{title}{Topological entanglement entropy}.
\newblock \emph{\bibinfo{journal}{Phys. Rev. Lett.}}
  \textbf{\bibinfo{volume}{96}}, \bibinfo{pages}{110404}
  (\bibinfo{year}{2006}).

\bibitem{Levin2006}
\bibinfo{author}{Levin, M.} \& \bibinfo{author}{Wen, X.-G.}
\newblock \bibinfo{title}{Detecting topological order in a ground state wave
  function}.
\newblock \emph{\bibinfo{journal}{Phys. Rev. Lett.}}
  \textbf{\bibinfo{volume}{96}}, \bibinfo{pages}{110405}
  (\bibinfo{year}{2006}).

\bibitem{Daley2012}
\bibinfo{author}{Daley, A.~J.}, \bibinfo{author}{Pichler, H.},
  \bibinfo{author}{Schachenmayer, J.} \& \bibinfo{author}{Zoller, P.}
\newblock \bibinfo{title}{Measuring entanglement growth in quench dynamics of
  bosons in an optical lattice}.
\newblock \emph{\bibinfo{journal}{Phys. Rev. Lett.}}
  \textbf{\bibinfo{volume}{109}}, \bibinfo{pages}{020505}
  (\bibinfo{year}{2012}).

\bibitem{Hahn}
\bibinfo{author}{Hahn, E.~L.}
\newblock \bibinfo{title}{Spin echoes}.
\newblock \emph{\bibinfo{journal}{Phys. Rev.}} \textbf{\bibinfo{volume}{80}},
  \bibinfo{pages}{580--594} (\bibinfo{year}{1950}).

\bibitem{Larkin1969}
\bibinfo{author}{Larkin, A.} \& \bibinfo{author}{Ovchinnikov, Y.~N.}
\newblock \bibinfo{title}{Quasiclassical method in the theory of
  superconductivity}.
\newblock \emph{\bibinfo{journal}{Sov. Phys. - JETP}}
  \textbf{\bibinfo{volume}{28}}, \bibinfo{pages}{1200} (\bibinfo{year}{1969}).

\bibitem{Fan2017}
\bibinfo{author}{Fan, R.}, \bibinfo{author}{Zhang, P.}, \bibinfo{author}{Shen,
  H.} \& \bibinfo{author}{Zhai, H.}
\newblock \bibinfo{title}{Out-of-time-order correlation for many-body
  localization}.
\newblock \emph{\bibinfo{journal}{Science Bulletin}}
  \textbf{\bibinfo{volume}{62}}, \bibinfo{pages}{707 -- 711}
  (\bibinfo{year}{2017}).

\bibitem{SwingleDisorderScrambling2017}
\bibinfo{author}{Swingle, B.} \& \bibinfo{author}{Chowdhury, D.}
\newblock \bibinfo{title}{Slow scrambling in disordered quantum systems}.
\newblock \emph{\bibinfo{journal}{Phys. Rev. B}} \textbf{\bibinfo{volume}{95}},
  \bibinfo{pages}{060201} (\bibinfo{year}{2017}).

\bibitem{Heyl_OTOCS_2018}
\bibinfo{author}{Heyl, M.}, \bibinfo{author}{Pollmann, F.} \&
  \bibinfo{author}{D\'ora, B.}
\newblock \bibinfo{title}{Detecting equilibrium and dynamical quantum phase
  transitions in Ising chains via out-of-time-ordered correlators}.
\newblock \emph{\bibinfo{journal}{Phys. Rev. Lett.}}
  \textbf{\bibinfo{volume}{121}}, \bibinfo{pages}{016801}
  (\bibinfo{year}{2018}).

\bibitem{Swingle_perspective_2018}
\bibinfo{author}{Swingle, B.}
\newblock \bibinfo{title}{Unscrambling the physics of out-of-time-order
  correlators}.
\newblock \emph{\bibinfo{journal}{Nature Physics}}
  \textbf{\bibinfo{volume}{14}}, \bibinfo{pages}{988--990}
  (\bibinfo{year}{2018}).

\bibitem{Rey_MQC_2017}
\bibinfo{author}{G{\"a}rttner, M.} \emph{et~al.}
\newblock \bibinfo{title}{Measuring out-of-time-order correlations and multiple
  quantum spectra in a trapped-ion quantum magnet}.
\newblock \emph{\bibinfo{journal}{Nature Physics}}
  \textbf{\bibinfo{volume}{13}}, \bibinfo{pages}{781--786} (\bibinfo{year}{2017}).

\bibitem{Du_ScramblingNMR_2017}
\bibinfo{author}{Li, J.} \emph{et~al.}
\newblock \bibinfo{title}{Measuring out-of-time-order correlators on a nuclear
  magnetic resonance quantum simulator}.
\newblock \emph{\bibinfo{journal}{Phys. Rev. X}} \textbf{\bibinfo{volume}{7}},
  \bibinfo{pages}{031011} (\bibinfo{year}{2017}).

\bibitem{Cappellaro2018}
\bibinfo{author}{Wei, K.~X.}, \bibinfo{author}{Ramanathan, C.} \&
  \bibinfo{author}{Cappellaro, P.}
\newblock \bibinfo{title}{Exploring localization in nuclear spin chains}.
\newblock \emph{\bibinfo{journal}{Phys. Rev. Lett.}}
  \textbf{\bibinfo{volume}{120}}, \bibinfo{pages}{070501}
  (\bibinfo{year}{2018}).

\bibitem{Bryce}
\bibinfo{author}{Meier, E.~J.}, \bibinfo{author}{Ang'ong'a, J.},
  \bibinfo{author}{An, F.~A.} \& \bibinfo{author}{Gadway, B.}
\newblock \bibinfo{title}{Exploring quantum signatures of chaos on a Floquet
  synthetic lattice}.
\newblock \emph{\bibinfo{journal}{arxiv}}
  \textbf{\bibinfo{volume}{1705.06714v1}} (\bibinfo{year}{2018}).

\bibitem{Monroe_Scrambling_2018}
\bibinfo{author}{Landsman, K.~A.} \emph{et~al.}
\newblock \bibinfo{title}{Verified quantum information scrambling}.
\newblock \emph{\bibinfo{journal}{Nature}} \textbf{\bibinfo{volume}{567}},
  \bibinfo{pages}{61--65} (\bibinfo{year}{2019}).
  
  

  
\bibitem{SYK}
\bibinfo{author}{Sachdev, S. and Ye, J.} \newblock \bibinfo{title}{Gapless spin-fluid ground state in a random quantum Heisenberg magnet}.
\newblock \emph{\bibinfo{journal}{Phys. Rev. Lett.}} \textbf{\bibinfo{volume}{70}},
  \bibinfo{pages}{3339--3342} (\bibinfo{year}{1993}).
  
  
  
  

\bibitem{Blatt_DigitalSimulation_2011}
\bibinfo{author}{Lanyon, B.~P.} \emph{et~al.}
\newblock \bibinfo{title}{Universal digital quantum simulation with trapped
  ions}.
\newblock \emph{\bibinfo{journal}{Science}} \textbf{\bibinfo{volume}{334}},
  \bibinfo{pages}{57--61} (\bibinfo{year}{2011}).
  
  
  
  \bibitem{Lucasgraph}
\bibinfo{author}{Lucas, A.} 
\newblock \bibinfo{title}{Quantum many-body dynamics on the star graph}.
\newblock \emph{\bibinfo{journal}{arxiv}} \textbf{\bibinfo{volume}{1903.01468}} (\bibinfo{year}{2019}).
  
  

\bibitem{Martinez2016}
\bibinfo{author}{Martinez, E.~A.} \emph{et~al.}
\newblock \bibinfo{title}{Real-time dynamics of lattice gauge theories with a
  few-qubit quantum computer}.
\newblock \emph{\bibinfo{journal}{Nature}} \textbf{\bibinfo{volume}{534}},
  \bibinfo{pages}{516} (\bibinfo{year}{2016}).

\bibitem{1203.5813}
\bibinfo{author}{Preskill, J.}
\newblock \bibinfo{title}{Quantum computing and the entanglement frontier}
  (\bibinfo{year}{2012}).
\newblock \eprint{arXiv:1203.5813}.

\bibitem{Keyserlingk2018}
\bibinfo{author}{von Keyserlingk, C.~W.}, \bibinfo{author}{Rakovszky, T.},
  \bibinfo{author}{Pollmann, F.} \& \bibinfo{author}{Sondhi, S.~L.}
\newblock \bibinfo{title}{Operator hydrodynamics, otocs, and entanglement
  growth in systems without conservation laws}.
\newblock \emph{\bibinfo{journal}{Phys. Rev. X}} \textbf{\bibinfo{volume}{8}},
  \bibinfo{pages}{021013} (\bibinfo{year}{2018}).

\bibitem{Khemani2018}
\bibinfo{author}{Khemani, V.}, \bibinfo{author}{Vishwanath, A.} \&
  \bibinfo{author}{Huse, D.~A.}
\newblock \bibinfo{title}{Operator spreading and the emergence of dissipative
  hydrodynamics under unitary evolution with conservation laws}.
\newblock \emph{\bibinfo{journal}{Phys. Rev. X}} \textbf{\bibinfo{volume}{8}},
  \bibinfo{pages}{031057} (\bibinfo{year}{2018}).

\bibitem{Nahum2018}
\bibinfo{author}{Nahum, A.}, \bibinfo{author}{Vijay, S.} \&
  \bibinfo{author}{Haah, J.}
\newblock \bibinfo{title}{Operator spreading in random unitary circuits}.
\newblock \emph{\bibinfo{journal}{Phys. Rev. X}} \textbf{\bibinfo{volume}{8}},
  \bibinfo{pages}{021014} (\bibinfo{year}{2018}).

\bibitem{Eckardt2017}
\bibinfo{author}{Eckardt, A.}
\newblock \bibinfo{title}{Colloquium: Atomic quantum gases in periodically
  driven optical lattices}.
\newblock \emph{\bibinfo{journal}{Rev. Mod. Phys.}}
  \textbf{\bibinfo{volume}{89}}, \bibinfo{pages}{011004}
  (\bibinfo{year}{2017}).

\bibitem{Zhang:2017ci}
\bibinfo{author}{Zhang, J.} \emph{et~al.}
\newblock \bibinfo{title}{{Observation of a discrete time crystal}}.
\newblock \emph{\bibinfo{journal}{Nature}} \textbf{\bibinfo{volume}{543}},
  \bibinfo{pages}{217--220} (\bibinfo{year}{2017}).

\bibitem{Choi2017}
\bibinfo{author}{Choi, S.} \emph{et~al.}
\newblock \bibinfo{title}{Observation of discrete time-crystalline order in a
  disordered dipolar many-body system}.
\newblock \emph{\bibinfo{journal}{Nature}} \textbf{\bibinfo{volume}{543}},
  \bibinfo{pages}{221--225} (\bibinfo{year}{2017}).

\bibitem{Deutsch:2010ky}
\bibinfo{author}{Deutsch, C.} \emph{et~al.}
\newblock \bibinfo{title}{Spin self-rephasing and very long coherence times in
  a trapped atomic ensemble}.
\newblock \emph{\bibinfo{journal}{Phys. Rev. Lett.}}
  \textbf{\bibinfo{volume}{105}}, \bibinfo{pages}{020401}
  (\bibinfo{year}{2010}).

\bibitem{Solaro:2016iv}
\bibinfo{author}{Solaro, C.} \emph{et~al.}
\newblock \bibinfo{title}{{Competition between Spin Echo and Spin
  Self-Rephasing in a Trapped Atom Interferometer}}.
\newblock \emph{\bibinfo{journal}{Phys. Rev. Lett.}}
  \textbf{\bibinfo{volume}{117}}, \bibinfo{pages}{163003}
  (\bibinfo{year}{2016}).

\bibitem{Piechon:2009cr}
\bibinfo{author}{Pi\'echon, F.}, \bibinfo{author}{Fuchs, J.~N.} \&
  \bibinfo{author}{Lalo\"e, F.}
\newblock \bibinfo{title}{Cumulative identical spin rotation effects in
  collisionless trapped atomic gases}.
\newblock \emph{\bibinfo{journal}{Phys. Rev. Lett.}}
  \textbf{\bibinfo{volume}{102}}, \bibinfo{pages}{215301}
  (\bibinfo{year}{2009}).

\bibitem{Baumann2010}
\bibinfo{author}{Baumann, K.}, \bibinfo{author}{Guerlin, C.},
  \bibinfo{author}{Brennecke, F.} \& \bibinfo{author}{Esslinger, T.}
\newblock \bibinfo{title}{Dicke quantum phase transition with a superfluid gas
  in an optical cavity}.
\newblock \emph{\bibinfo{journal}{Nature}} \textbf{\bibinfo{volume}{464}},
  \bibinfo{pages}{1301--1306} (\bibinfo{year}{2010}).

\bibitem{Klinder2015}
\bibinfo{author}{Klinder, J.}, \bibinfo{author}{Ke{\ss}ler, H.},
  \bibinfo{author}{Wolke, M.}, \bibinfo{author}{Mathey, L.} \&
  \bibinfo{author}{Hemmerich, A.}
\newblock \bibinfo{title}{Dynamical phase transition in the open Dicke model}.
\newblock \emph{\bibinfo{journal}{PNAS}} \textbf{\bibinfo{volume}{112}},
  \bibinfo{pages}{3290--3295} (\bibinfo{year}{2015}).

\bibitem{Leonard2017bis}
\bibinfo{author}{Leonard, J.}, \bibinfo{author}{Morales, A.},
  \bibinfo{author}{Zupancic, P.}, \bibinfo{author}{Donner, T.} \&
  \bibinfo{author}{Esslinger, T.}
\newblock \bibinfo{title}{Monitoring and manipulating Higgs and Goldstone modes
  in a supersolid quantum gas}.
\newblock \emph{\bibinfo{journal}{Science}} \textbf{\bibinfo{volume}{358}},
  \bibinfo{pages}{1415--1418} (\bibinfo{year}{2017}).

\bibitem{Li:2017}
\bibinfo{author}{Li, J.} \emph{et~al.}
\newblock \bibinfo{title}{{A stripe phase with supersolid properties in spin
  orbit-coupled Bose-Einstein condensates}}.
\newblock \emph{\bibinfo{journal}{Nature}} \textbf{\bibinfo{volume}{543}},
  \bibinfo{pages}{91--94} (\bibinfo{year}{2017}).

\bibitem{Monroe_53Ion_2017}
\bibinfo{author}{Zhang, J.} \emph{et~al.}
\newblock \bibinfo{title}{Observation of a many-body dynamical phase transition
  with a 53-qubit quantum simulator}.
\newblock \emph{\bibinfo{journal}{Nature}} \textbf{\bibinfo{volume}{551}},
  \bibinfo{pages}{601--604} (\bibinfo{year}{2017}).

\bibitem{Jurcevic2017}
\bibinfo{author}{Jurcevic, P.} \emph{et~al.}
\newblock \bibinfo{title}{Direct observation of dynamical quantum phase
  transitions in an interacting many-body system}.
\newblock \emph{\bibinfo{journal}{Phys. Rev. Lett.}}
  \textbf{\bibinfo{volume}{119}}, \bibinfo{pages}{080501}
  (\bibinfo{year}{2017}).

\bibitem{Smale2018}
\bibinfo{author}{Smale, S.} \emph{et~al.}
\newblock \bibinfo{title}{Observation of a dynamical phase transition in a
  quantum degenerate Fermi gas}.
\newblock \emph{\bibinfo{journal}{arXiv:1806.11044}}  (\bibinfo{year}{2018}).

\end{thebibliography}


%
%

\end{document}